# NeuroQuantify – An Image Analysis Software for Detection and Quantification of Neurons and Neurites using Deep Learning

**Ka My Dang**[1,2*], **Yi Jia Zhang**[3], **Tianchen Zhang**[3], **Chao Wang**[3], **Anton Sinner**[1], **Piero Coronica**[4], **and Joyce K. S. Poon**[1,2,3*]

[1]*Max Planck Institute of Microstructure Physics, Weinberg 2, D-06120 Halle, Germany*

[2]*Max Planck-University of Toronto Centre for Neural Science and Technology, Canada*

[3]*Department of Electrical and Computer Engineering, University of Toronto, 10 King's College Rd., Toronto, Ontario M5S 3G4, Canada*

[4]*Max Planck Computing and Data Facility, Gießenbachstraße 2, 85748 Garching, Germany*

Correspondence*:

Ka My Dang

kamydang@mpi-halle.mpg.de

Joyce K. S. Poon

joyce.poon@mpi-halle.mpg.de

**Abstract:**

The segmentation of cells and neurites in microscopy images of neuronal networks provides valuable quantitative information about neuron growth and neuronal differentiation, including the number of cells, neurites, neurite length and neurite orientation. This information is essential for assessing the development of neuronal networks in response to extracellular stimuli, which is useful for studying neuronal structures, for example, the study of neurodegenerative diseases and pharmaceuticals. However, automatic and accurate analysis of neuronal structures from phase contrast images has remained challenging. To address this, we have developed NeuroQuantify, an open-source software that uses deep learning to efficiently and quickly segment cells and neurites in phase contrast microscopy images. NeuroQuantify offers several key features: (i) automatic detection of cells and neurites; (ii) post-processing of the images for the quantitative neurite length measurement based on segmentation of phase contrast microscopy images, and (iii) identification of neurite orientations. The user-friendly NeuroQuantify software can be installed and freely downloaded from GitHub https://github.com/StanleyZ0528/neural-image-segmentation.

## 1.   Introduction

Quantitative analysis of neuronal cell structures is important for biomedical and pharmaceutical research, such as the determination of drug uptake and toxicity[1,2]. Typical analysis involves monitoring changes in the culture properties, such as the neuron numbers, neurite outgrowth directions, and neurite differentiation, to assess the physiological state of the neuronal culture[3]. Changes in the neuronal networks are indicative of neuronal development in response to extracellular stimuli (e.g., biochemical, electrical, optical, mechanical, and topographical)[4–8], and properties such as cell numbers and neurite lengths can serve as cues for such changes[9,10]. For instance, blue light exposure can cause the retraction of neurites in neurons differentiated from neuroblastoma cells, resembling pathological neurite degradation, while red light can induce the regrowth of retracted neurites[11,12]. Furthermore, the direction of neurite extension provides insights into neurite outgrowth and nerve guidance[13,14]. However, the analysis of phase contrast biological images is challenging due to the presence of halo and shade-off artifacts[15], as well as the diverse shapes and sizes of neurons, making segmentation difficult[16]. In addition, measuring the neurite length and direction of neurite extension typically involves manual tracing, which is time-consuming and may yield inconsistent results in repeated measurements. To address these challenges, numerous image-processing algorithms have been developed using software packages such as ImageJ and toolboxes in Matlab, enabling semi-automatic or automatic detection and quantification of neuronal structures[17–20]. The most commonly used algorithms for analyzing neuronal development are catered to fluorescence microscopy that utilize indicators to color cells and neurites. However, depending on the application, modifying cells for fluorescence microscopy is not always possible[21,22].

Recently, supervised learning using deep learning has offered a solution to analyzing phase-contrast microscope images that overcomes the limitations of conventional methods[23]. Neural networks, specifically convolutional neural networks (CNNs), have shown success in cell segmentation[24], providing more accurate segmentations with greater robustness[25]. Among CNN-based methods, U-Net has emerged as the most widely adopted approach for image segmentation, delivering promising results in live cell images[26,27]. However, manually creating image segmentation masks for training models is a time-intensive process, resulting in a limited number of training images[28]. Furthermore, deep learning approaches for quantitative biological images have mostly focused on cell morphology[29,30], or single-cell segmentation among multiple types of cells[31]. The automated segmentation of cells and neurites in phase-contrast microscopy images, which lack the color differentiation in fluorescence microscopy, using deep learning models remains challenging, likely due to the thin and complex physical structures of the neurites. An effective model is needed to achieve high accuracy in neuronal structure segmentation with a limited number of labeled images.

In this paper, we develop a well-tuned machine learning model for neuronal image segmentation based on a modified U-Net architecture. Additionally, we present a software package called NeuroQuantify, which offers functionalities such as cell and neurite detection, counting, neurite length measurement, and neurite orientation distribution. This comprehensive tool enables quick and efficient quantitative evaluation of neuronal circuits, providing valuable insights into neuronal networks on a large scale. NeuroQuantify is implemented in Python 3 using open-source packages and is freely available for download and local installation from GitHub. Its user-friendly graphical interface facilitates precise annotation of cells and neurites from phase-contrast microscopy images, making it an invaluable resource for investigating biological questions concerning neuronal networks.

## 2. Material and Methods:

We introduce a neuron quantification method based on deep learning for cell and neurite segmentation. Our method uses phase-contrast microscope images and labeled images as masks to train the neural network. After image segmentation, an algorithm of image post-processing is performed for neuron quantification. **Figure 1** illustrates our computational pipeline: First, a deep learning model classifies features in the image as cells or neurites (**Figure 1a),** then the cells and neurites are counted, and the lengths and orientations of the neurites are measured **(Figure 1b)**.

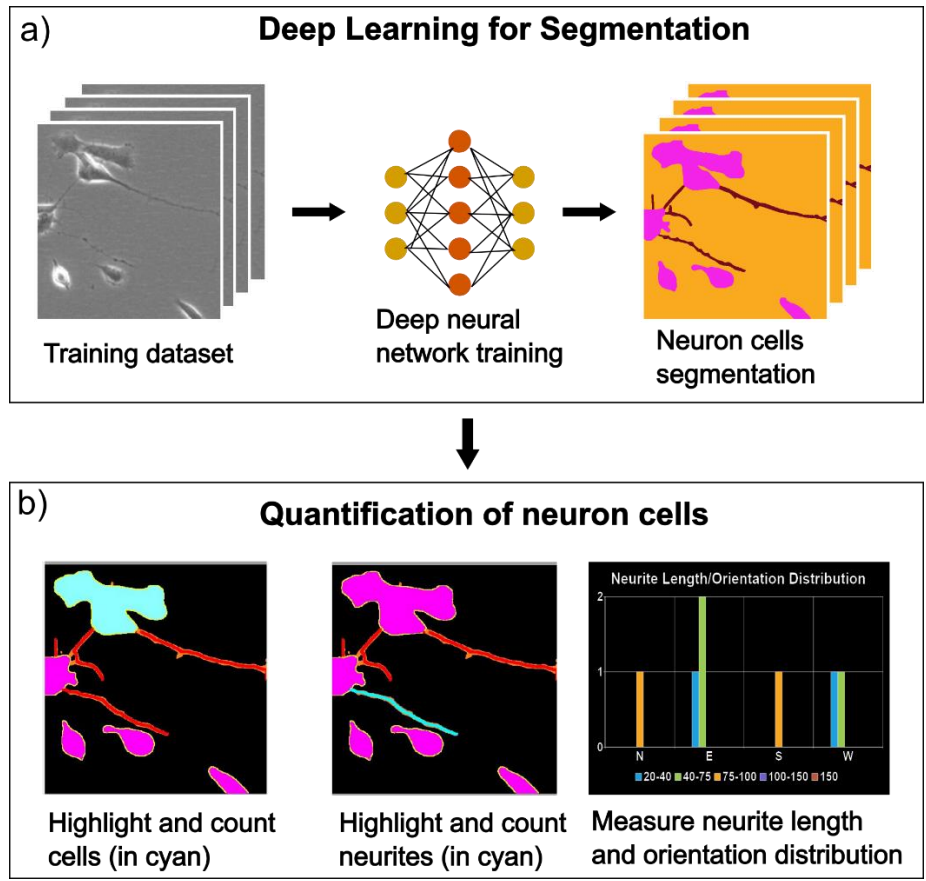

**Figure 1**: Overview of the computational pipeline of NeuroQuantify, a) Deep learning for cell and neurite segmentation, and b) Quantification number of cells and measurement of neurite lengths and its orientation distribution.

## 2.1. Dataset
### *Data Acquisition and Preprocessing Methodology*
*Dataset collection:* The dataset employed in this study encompasses two-dimensional (2D) phase-contrast microscopy images of neuroblastoma cells (SH-SY5Y). These neuron-like cells were grown in a T25 flask and treated with Retinoic acid (R2625, Sigma-Aldrich) following a standard protocol[32]. The imaging process was initiated on the third day of cultivation, using a Zeiss microscope with a 10x magnification objective and phase contrast mode. Multiple regions of interest were selected on each T25 flask, focusing on the areas with a high density of cells and neurites. Each image has a resolution of 2560×1920 pixels.

*Manual Annotation and Class Labeling:*
To facilitate subsequent analysis, a total of 200 images (2563×1920 pixels) were manually annotated, using a specialized software provided by ByteBridge. Three different classes on the image were assigned distinct colors corresponding to cells, neurites, and background.

*Image enhancement through gamma correction:*
The raw phase-contrast microscope images displayed variations in background brightness levels, which could potentially introduce inconsistencies during subsequent processing. To ensure consistency in the output images, gamma correction is applied according to the following equations:

$$\gamma = \frac{log\ (255 \times 0.5)}{log\ (mean(Input\ image\ gray\ values))}, \quad (1a)$$
$$Output\ image = (Input\ image)^\gamma. \quad (1b)$$

From Eq. 1a, the original image brightness value is first compared to the relative brightness parameter $log(255 \times 0.5)$, and then the adjustment is conducted using Eq. 1b. This correction process equalizes the brightness level, reducing bias in the subsequent training process.

*Image Cropping and Dataset Generation:* The initial phase-contrast images, with a size of 2563×1920 pixels, were divided into training, validation, and test sets. Each of these images was cropped into 20 smaller images each with a smaller size of 512×512 pixels (**Figure 2**). During the crop, filtering was applied to remove the small images (512×512 pixels) that contained a scale bar on the corner of the image and those displaying mostly background. The 512×512 size was selected due to the memory limitation during the model training process. The final datasets consist of training dataset (2740 frames), validation set (247 frames), and test set A (323 frames), where the size of each frame was 512×512 pixels. To accurately assess cellular and neurite counts, as well as neurite lengths within the high-resolution images (2563×1920 pixels), an additional test set of 20 images with 2563×1920 pixels was deliberately introduced. This supplementary dataset, called test set B, served the explicit purpose of evaluating the performance of the post-processing phase, i.e. the analysis of the segmentation masks, and has not been used during the training of the neural network. The segmentation mask of these high-resolution images has been enriched by information on the number of cells and neurites manually counted by an expert using ImageJ, and the average length of the neurites as given by the NeuronJ plugin.

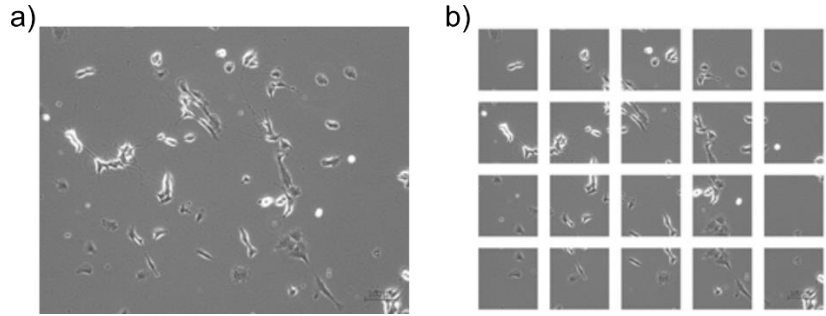

**Figure 2:** a) An original image, with a size of 2653×1920, and b) 20 smaller images with a size of 512×512 from the original image.

**Data availability:** The datasets generated and analyzed during the current study are available in the Edmond of the Max Planck society repository.
https://edmond.mpdl.mpg.de/dataset.xhtml?persistentId=doi:10.17617/3.UIBMJX

## 2.2. Neural Network Architecture:

### U-Net for Semantic Segmentation

The models used in the paper are based on U-Net, which consists of an encoder and a decoder. U-Net is a CNN specifically designed for biomedical image segmentation[22]. Its architecture includes a contracting path to capture context and an expanding path for precise localization. The model generates a pixel-by-pixel mask that represents the class of each pixel. One major advantage of the U-Net model is its ability to learn effectively from a relatively small dataset.

**Figure 3** illustrates the architecture diagram for the primary model used in our work, referred to as the large model. This model consists of 4 down-sampling blocks and 4 up-sampling blocks, with an initial convolutional block with 64 output channels. In each down-sampling block, the input data undergoes two consecutive 3×3 kernel size convolutions with a ReLu activation function. Subsequently, a 2×2 kernel max pool layer is applied to reduce data size. After the fourth block, the data enters the up-sampling path. During up-sampling, the data undergoes a reverse convolution layer with a 2×2 kernel and half the original number of features. It is then concatenated with a copy of the data outputted at the same block level in the down-sampling path. The combined tensor is passed through a double convolution layer, with the number of output channels matching the reverse convolution step. Finally, after 4 blocks of up-sampling, the data goes through a 3×3 convolutional layer with an output channel size set of 3, performing the final three-class segmentation task. In addition to the large model, there is also a simplified model, named small model. This model comprises 3 blocks of down-sampling and 3 blocks of up-sampling. The number of output channels in the initial convolution block is reduced to 16 features, significantly reducing the random access memory usage during training. The small model demonstrates more efficient computation in practice.

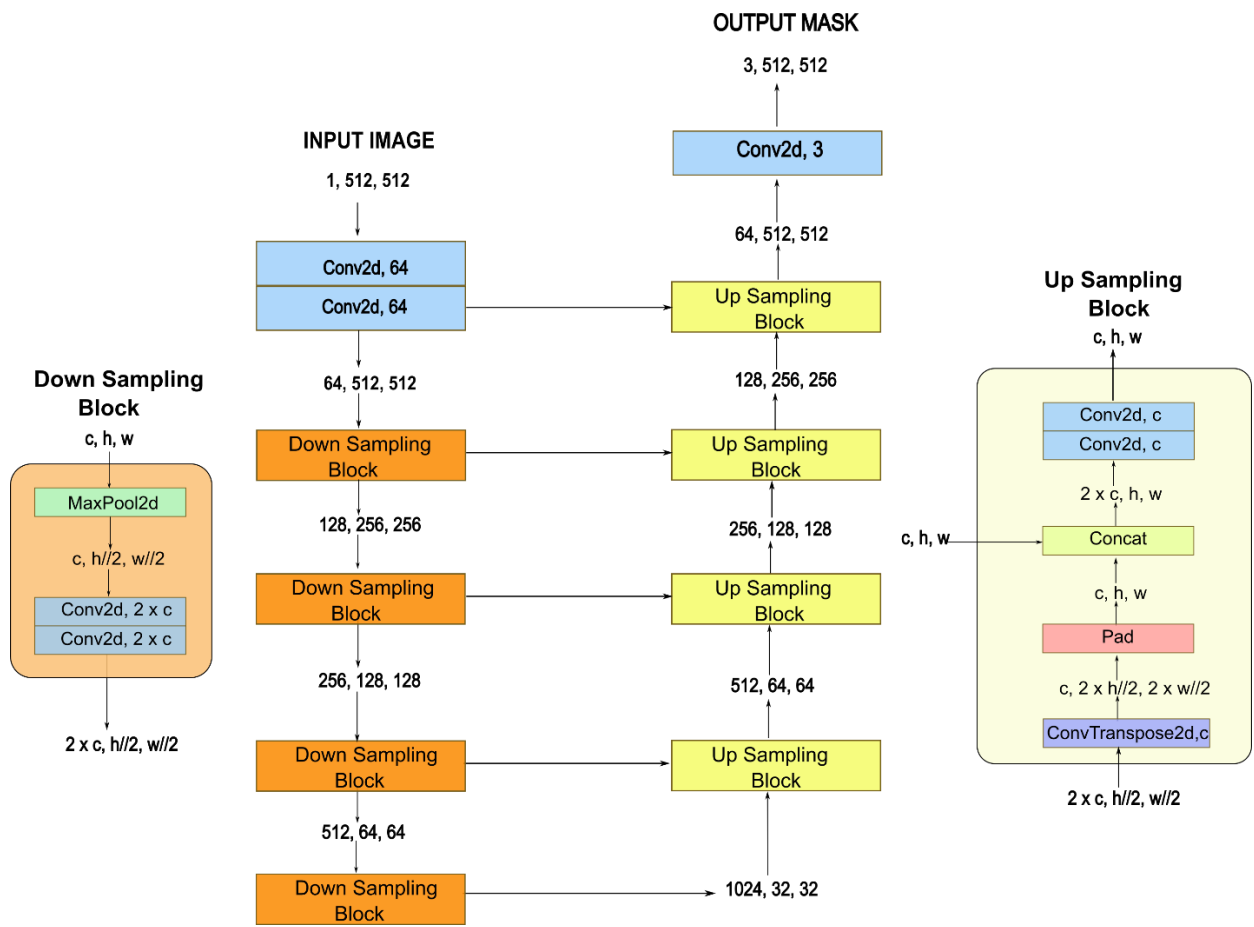

**Figure 3:** U-Net architecture in NeuroQuantify (large model). The large model uses 4 down sampling blocks and 4 up sampling blocks, while only 3 of each are used for small model. Both models use ReLU as activation functions. Down Sampling blocks extract and transform features, simultaneously reducing the spatial dimensions. Up sampling block restore spatial dimensions for improved localization. In the diagram, the shape of the tensors is also shown in channel-first format (c, h, w representing channels, the height and the width of a tensor).

Training process:

The models have been trained on the 2740 512x512 frames in the training dataset. For both the large and small models, we used the Adam optimizer with a learning rate of 0.0001. The learning rate, as well as hyper-parameters, was selected through a process of optimization, driven by the generalization performance of our models evaluated on the 247 images in the validation set.

Loss function

We used a combination of cross entropy loss (CE)[33] and soft dice loss (D)[34] as the loss function to train both the small and large models. CE is widely used for classification tasks, and it measures the dissimilarity between the predicted probability distribution and the ground truth label for each pixel. To emphasize the segmentation performance for the cells and neurites while minimizing the impact on the background pixels, rescaling weight were assigned to each class when computing the overall CE. Specifically, the weight factor was set to 2.0 for background, 3.0 for cell bodies, and 5.0 for neurites.

The D function is computed individually for each class separately and then averaged to obtain a final score[35]. Using soft dice loss, the model mitigates the issue of inflated accuracy caused by correctly classifying the background, which takes most of the region within the image. Instead, it prioritizes the accuracy of cell classification by assessing intersectional aspects.

The global loss functions of the large and small models are, respectively,

$$L_{large} = 0.5D + 0.5CE, \qquad (2a)$$
$$L_{small} = 0.8D + 0.2CE. \qquad (2b)$$

## 2.3. Image Post Processing:

*Algorithm for Cell Counting and Neurites Analysis from Segmentation Mask*

After performing image segmentation, we proceeded to group non-adjacent cells into cell clusters, while non-adjacent neurites were grouped into clusters based on their connections. Subsequently, we assigned the corresponding neurite groups to their respective cell clusters. We considered neurites originating from cells within the image while excluding those originating from cells outside the image.

To count the number of cells, we started by establishing a typical area based on the histogram of the cell area distribution on 200 images of our training dataset (**Figure S1**). We observed that the distribution centered around approximately 40µm in diameter, with the corresponding cell area of approximately 1256 µm$^2$. For each cell cluster, we calculated the number of cells by dividing the cluster area by the typical cell area and rounding to the nearest integer. Adding the number of cells across all clusters yielded the final cell count for the image.

To determine the number of neurites and analyze their length, we used the segmentation mask. The segmentation mask categorizes each pixel as background, cells or neurites accordingly. **Figure 4a** shows the segmentation mask with three colors for background, cells and neurites. The masked pixels for the cells and neurites are then filtered out separately by colors as shown in **Figure 4b** and **4c.** To convert the segmentation neurite region to a single-pixel-width skeleton along the center region, we used skeletonization in the fil-finder library[36], resulting in the neurite skeleton shown in **Figure 4d**. In this figure, white lines represent the neurite skeleton, green dots indicate "end points"; blue dots signify "intersection points", and red dots denote "touch points". The presence of "touch points" aims to include the connectivity of the neurites and cells in the analysis. They are the closest points on the skeleton where the neurite and the cell are connected. The "touch points", "intersection points" and "end points" separate the neurites into smaller branch segments and do further analysis. After the detection of neurites, NeuroQuantify measures the length and orientation of the detected neurites. The length of a neurite is calculated based on the scale information of captured images (1μm = 2.21 pixel). We only count neurites with lengths longer than 20 μm to eliminate short neurites that are insignificant for the analysis and potential noise from the segmentation mask. In case where multiple neurites protrude from a single location in the cluster, we only count the longest. When the neurites cross each other, we trace their origins. Therefore, the algorithm is designed to find a "touching" branch segment which starts from at least one "touch point" and use it as the starting of a branch. Then we assign additional branch segments that are directly connected to the previous branch. This process is repeated until no branch segment can be found. In case there are multiple branches connected to the previous branch, the program counts the extending branch with the longer projection length on the existing branch. The longer projection branch is illustrated as **Figure 4e.** By applying these criteria, the data can be effectively analyzed and processed to obtain meaningful insights.

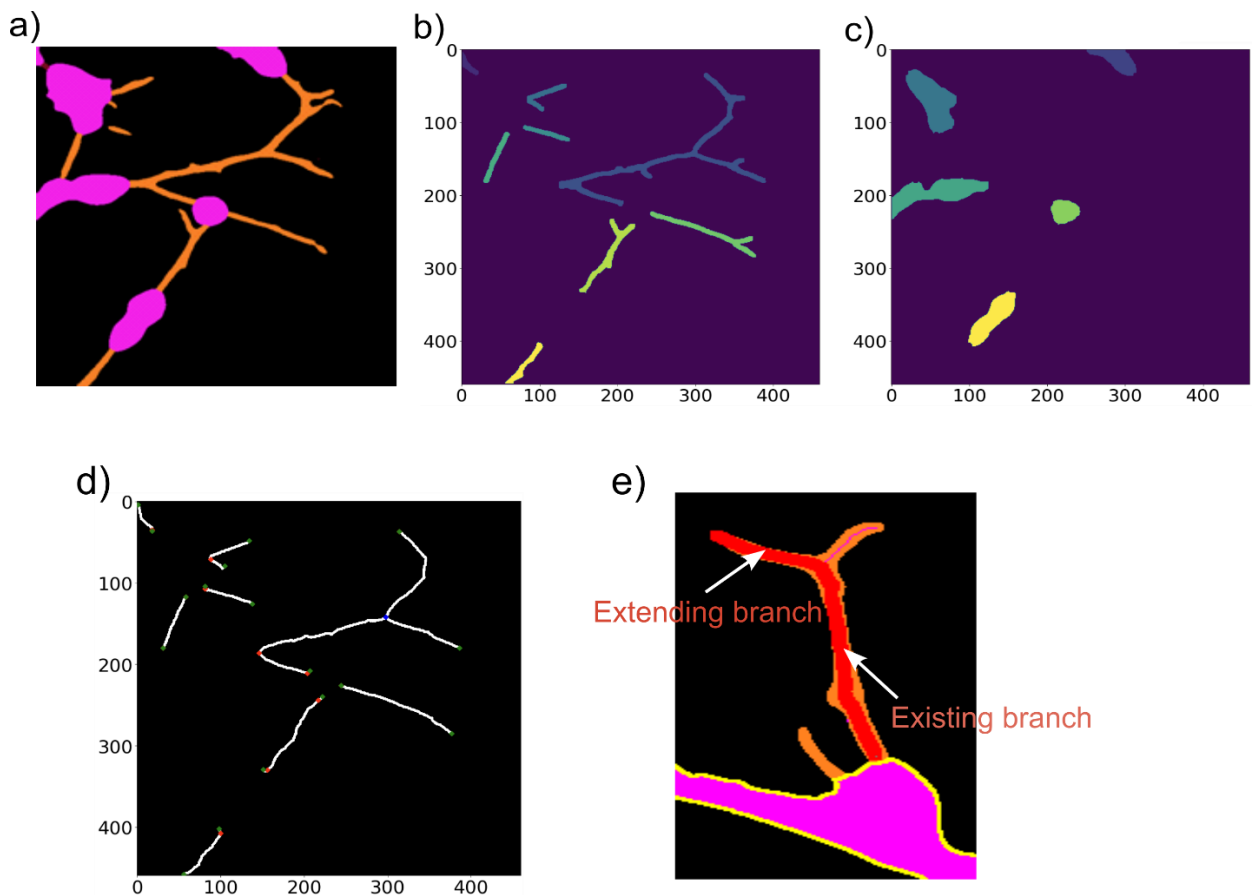

**Figure 4:** Post-image processing for neurite length quantification a) segmentation of prediction image, b) filtered image of neurites, c) filtered image of cells with different colors, d) neurite skeleton, e) priority of adding an extending branch into an existing branch.

### 2.4. Evaluation metrics:

*Metrics for semantic segmentation: Precision, Recall, F1 score, IoU score*

To evaluate the performance of our deep learning models, we computed the metrics including total precision, per-class recall, and F1 score. The metrics are defined as follows,

$$Total\ Precision = \frac{TP_{cell}+TP_{neurite}+TP_{background}}{TP_{cell}+TP_{axon}+TP_{bg}+FP_{cell}+FP_{neurite}+FP_{background}}, \quad (3a)$$

$$Recall = \frac{TP}{TP+FN}, \quad (3b)$$

$$F1\ score = \frac{2 \times Precision \times Recall}{Precision+Recall}, \quad (3c)$$

where TP, FP, and FN are respectively per class true positive, false positive, false negative. The detailed definition of TP, FP and FN for multi-classification can be seen in the supplementary.

The intersection over union (IoU) score is computed to estimate how well the segmentation of each class matches the ground truth mask at the pixel level. The IoU score is given by

$$IoU = \frac{A \cap B}{A \cup B}, \quad (3d)$$

where *A* is a predicted class and *B* is the corresponding ground truth mask.

*Metrics for Cell, Neurite counting and Average neurite length Accuracy*

The accuracy is calculated by comparing the predicted cell, neurite count and average neurite length from the predicted segmentation images analyzed by NeuroQuantify with the ground truth images. The equation is as follows:

$$Accuracy = 1 - \frac{|Predicted - GroundTruth|}{GroundTruth}. \quad (4)$$

## 3. Results

In the sections to follow, we summarize the results of the model. We begin by evaluating the large and small models in the segmentation task using the test set A in section 3.1. In section 3.2, we present the analysis pipeline including the post processing phase. We discuss the accuracy in the cells and neurites counting tasks by comparing the prediction of NeuroQuantify with the annotations by experts, and the neurite length measurement using NeuroQuantify is compared with the NeuronJ plugin from ImageJ. In section 3.3, we highlight the advantages of our post-image processing, including cell counting with neurites, neurite orientation distribution. Finally, section 3.4 introduces a user-friendly graphical interface and provides information regarding its processing time for two models.

### 3.1. Validating the segmentation performance in the large and small models

We evaluated the performances in the segmentation task of two deep learning models using the images of neuroblastoma cells contained in test set A. As discussed in section 2.4, to quantitatively evaluate the quality of a segmentation mask we used the total precision, the per-class recall, F1 score, and the IoU score (Eq. 3a, 3b, 3c, and 3d)

To assess the statistical significance of performance between the two models, we employed the two-sided Wilcoxon signed-rank test with a significance level of 0.05. **Figure 5** shows the results. The average total precision for three classes of segmentation in both models is very high, reaching approximately 0.98. **Figure 5b** shows that the large model exhibits significantly higher average recall values for cells and neurites compared to the small model (0.98 vs. 0.93 for cells, and 0.95 vs. 0.75 for neurites), suggesting that the large model more effectively detects cells and neurites. There is a significant difference in the F1 score for the cell and neurite class between the two models as shown in **Figure 5c**.

To further evaluate the precision of object detection in both models, we used the IoU score, which measures the overlap between the predicted image and the ground truth mask. **Figure 5d** illustrates the result. Both models attain a high IoU matching, with the large model achieving an IoU score of 0.84, while the small model achieves a score of 0.81. **Figure 5e** presents the confusion matrices for both the large and small models, highlighting the improved accuracy in cells and especially the neurite detection achieved by the large model.

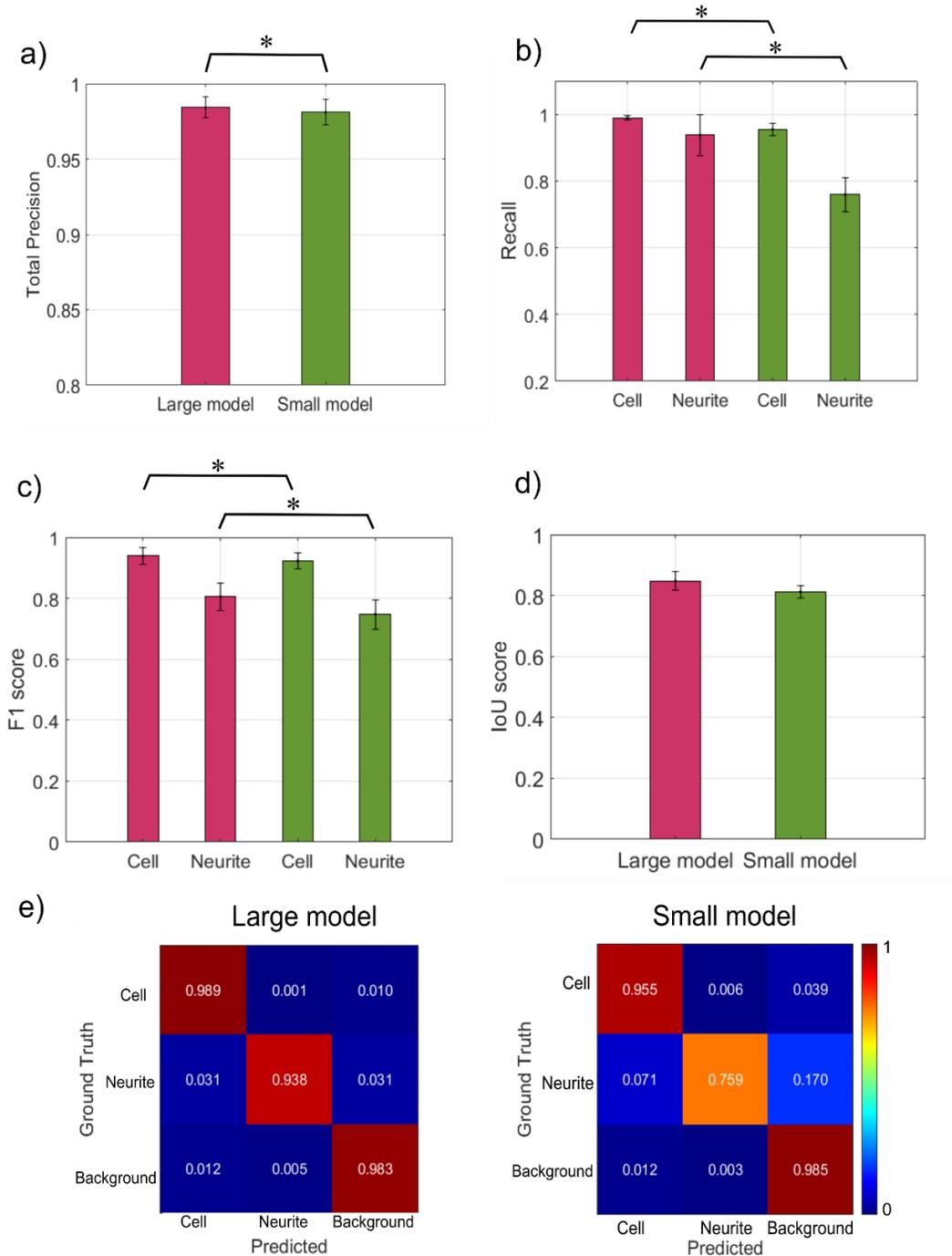

**Figure 5:** Comparison of performance for two models in NeuroQuantify. Significance was tested by the two-sided Wilcoxon signed-rank test, (*): $p < 0.05$; The bar plots shows the average and the standard deviation across the images in test set A of a) the total precision; b) the cell and neurite recall; c) the cell

and neurite F1 score, d) the IoU score, and e) confusion matrices of the segmentation performance for the large and small models.

### 3.2. Assessing Cell and Neurite Count Accuracy with NeuroQuantify

*Accuracy of Cell and Neurite counting using NeuroQuantify*

We used test set B consisting of 20 high-resolution phase-contrast images (2563x1920 pixels) to assess the capabilities of NeuroQuantify in counting cells and neurites by comparing the predicted images from NeuroQuantify with the ground truth. In **Figure 6a**, we present the accuracy of cell and neurite classification. Interestingly, the small model exhibits slightly higher accuracy in cell prediction compared to the large model. Specifically, the small model achieves an accuracy of 0.83, while the large model achieves 0.79. Regarding neurite detection, the small model achieves 0.65 accuracy, while the large model achieves 0.77.

We examined the inaccurate segmentation of the small model for neurite detection by investigating the predicted masks (segmented image from the model), where neurites are colored in orange, and the skeleton masks (image after post-processing), where neurites are colored in red. As shown in **Figure S2a-d**, which provides examples of neurite detection, both the large and small models exhibited a similar ability to detect long neurites (with lengths > 22 μm). However, when using the skeleton mask for neurite counting, the small model tended to overlook short neurites (with lengths < 22 μm). This issue is further illustrated in **Figure S2e-g**, where the small model failed to detect the short neurites.

*Accuracy of neurite length quantification using NeuroQuantify*

A feature of NeuroQuantify is the quantification of the neurite length. To assess the effectiveness of NeuroQuantify in measuring neurite lengths, we compared the average results on test set B with the ground truth obtained by NeuronJ, an ImageJ plugin[37]. To determine the statistical significance between the two methods, we conducted a two-sided Wilcoxon signed-rank test at a significance level of 0.05. **Figure 6b** illustrates the comparison of average neurite length measurements between NeuroQuantify and manual measurement. The results reveal that there is no significant difference between the manual technique and the large model ($p > 0.05$), whereas the small model exhibits a significant difference in neurite length measurement ($p < 0.05$). Furthermore, **Figure 6c** presents a plot illustrating the average neurite length accuracy for test set B. As depicted in **Figure 6c**, the average neurite length accuracy of both models is quite similar, with a value of 0.88 for the large model, and 0.84 for the small model.

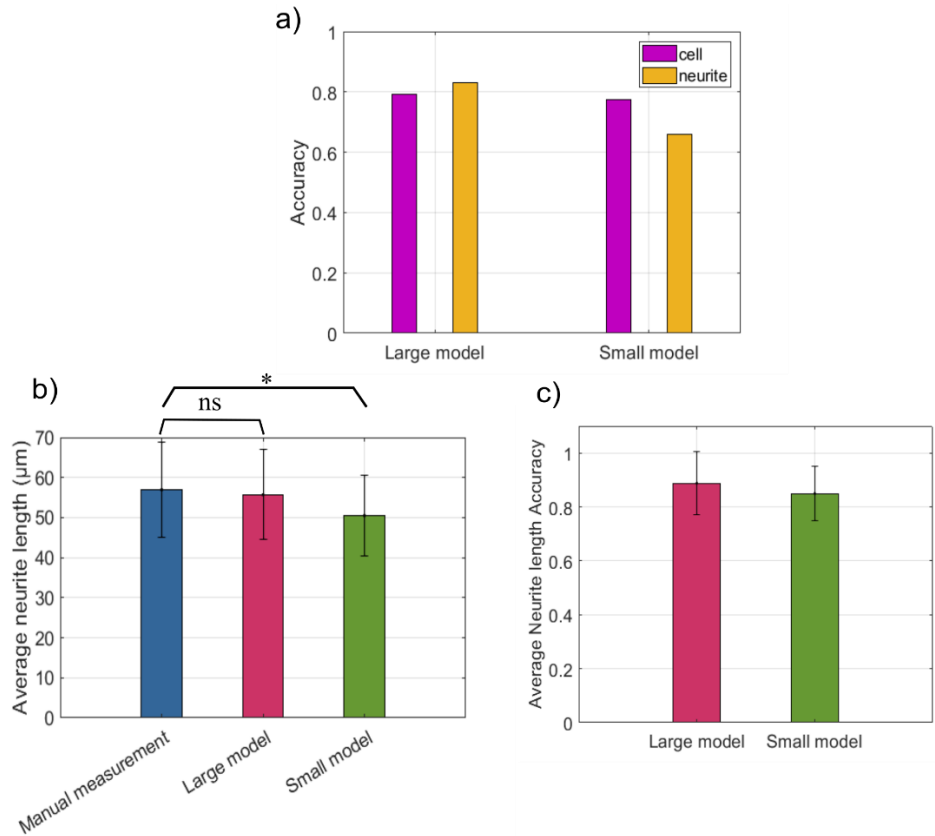

**Figure 6:** Evaluation of the segmentation performance of NeuroQuantify in the small and large models by calculating the accuracy of number of cells, neurites and neurite length quantification, Significance was tested by the two-sided Wilcoxon signed-rank test, ns: p > 0.05; *: p < 0.05; a) the accuracy of the number of cells and number of neurites as predicted by both models, b) a comparison between manual measurement and automated measurement using the large and small models, and c) a plot of average and standard deviation of neurite length accuracy for both models.

### 3.3. Neurites orientation distributions and counting cells with neurites using NeuroQuantify

NeuroQuantify offers an additional feature of providing the orientation distribution of neurites in the images. The orientation of a neurite is determined by drawing a straight line from the point where it attaches to the cell (known as the touch point) to its endpoint. If the other end of the neurite is connected to another cell, the orientation will be displayed in both directions. Four directions, namely north (N), south (S), east (E), and west (W), are defined to represent the neurite orientation. **Figure 7a** and **7b** illustrate the post-image processing steps involved in analyzing the neurite orientation. **Figure 7c** and **7d** provide an example of a single cell with three neurites, each pointing in a different direction.

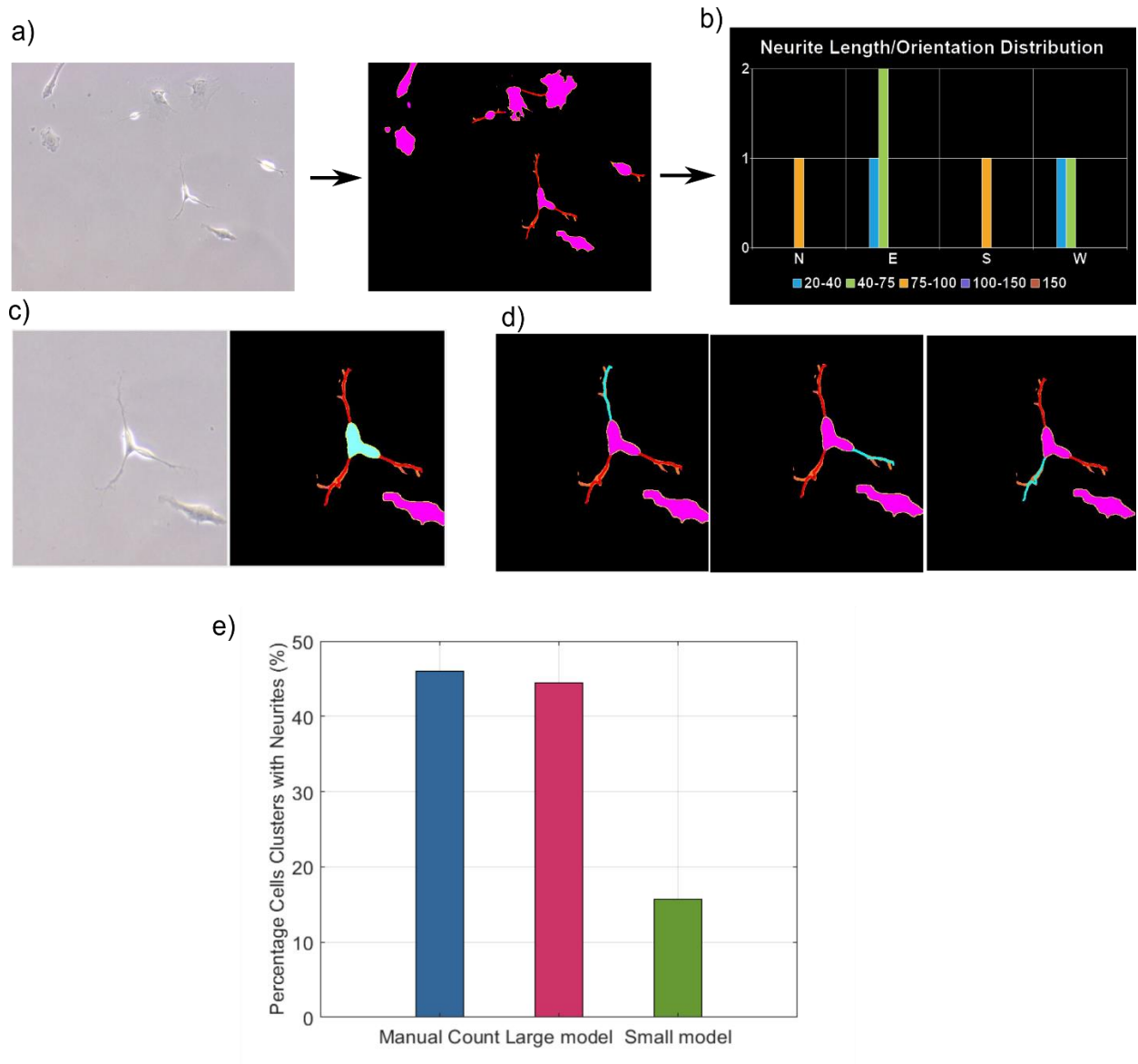

**Figure 7:** Post-processing of the images for analyzing neurite orientation distribution. a) an example of image post-processing from the phase-contrast image input to annotation mask and b) a plot of neurite length and neurite orientation distribution, c) and d) show an example of definition of neurite orientation distribution, and e) Percentage of cell clusters with neurite counted manually and using the large and small models

Furthermore, NeuroQuantify offers the capability to count the number of cells that have neurites. The algorithm specifically counts cells with neurites attached to them based on an adjacency analysis of the skeleton mask. The comparison of the manual counting and performance of the NeuroQuantify models in terms of the average percentage of cells with neurites is shown in **Figure 7e**. As shown in the figure, the small model was less effective in recognizing neurites when connected to cell clusters. Conversely, the large model demonstrates a similar percentage of cell clusters with neurites compared to manual counting.

## 3.4. NeuroQuantify User Interface and Processing time

To make the NeuroQuantify easier to use, we have developed a graphical user interface, as shown in **Figure 8a**. The user interface utilizes the Python library PyQt6 as its primary framework. This choice of framework ensures cross-platform compatibility, enabling users to access NeuroQuantify seamlessly on Windows 10 (1809 or later), MAC OS (12/11/10.15) (64bit Intel, 64bit ARM; XCode 12), and Linux (Ubuntu 20.04 (64bit Intel; gcc9), CentOS Linux 8.2, SLES 15 SP2 (SUSE Linux Enterprise Server, 64bit Intel; gcc10), Open SUSE 15.3 (64bit; gcc9)). Additionally, since the main algorithm and machine learning model have been implemented in Python, the integration of PyQt6 provides a more streamlined connection for Python-based functionalities. Within this user interface, as illustrated in **Figure 8a**, users can interact with NeuroQuantify by simply clicking on cells or neurites of interest. Upon selection, NeuroQuantify highlights the selected elements and provides relevant information. Moreover, users have the option to export neurite length information as a csv file, while the generated plot depicting neurite orientation and annotated data is automatically saved in the result folder. The analysis of neurite orientation offers valuable insights for studying axon guidance in diverse local environments[38,39].

To assess the performance of NeuroQuantify, we compared the processing time between two models using 20 images in the test set B. The processing time encompasses segmentation, annotation and post-image processing, which involves tasks such as cell and neurite counting, neurite length measurement, and neurite orientation distribution analysis. This analysis was conducted on a laptop with an 11[th] generation Intel i7-11800H CPU, 16GB of RAM, Windows 10. **Figure 8b** presents the processing time for both models. As depicted in **Figure 8b**, the processing time for the large model is approximately five times longer than that of the small model.

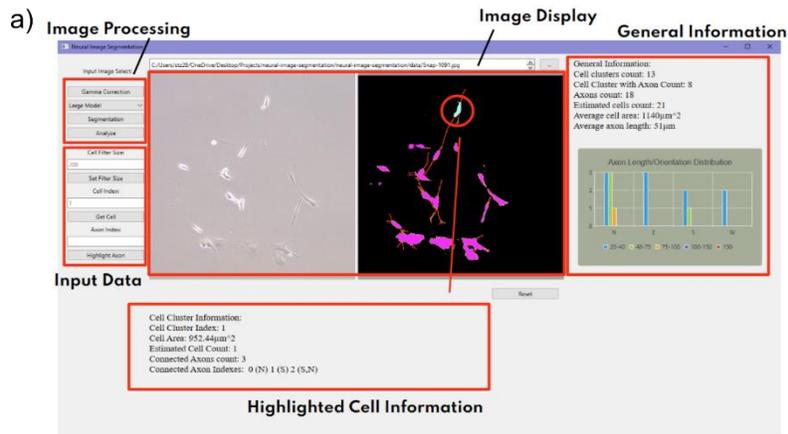

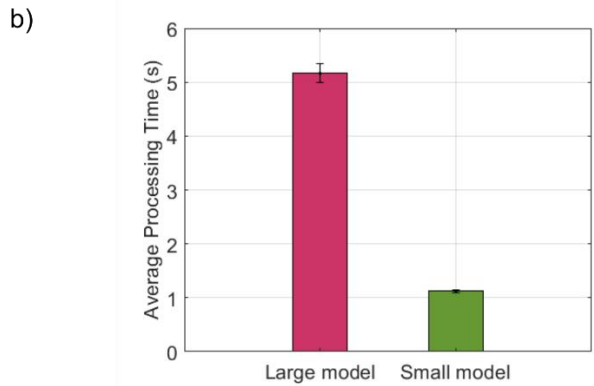

**Figure 8:** a) User Interface of NeuroQuantify and b) a comparison of the average processing time of the large and small models of NeuroQuantify

## 4. Discussion

We have introduced a comprehensive framework for neuron detection and semantic segmentation in images of neuron-like neuroblastoma cells. Automatic quantification of phase-contrast neuron culture images can accelerate laboratory investigations, and neurites are studied in the context of neuron regeneration and neurodegenerative diseases[40–43]. Neuroblastoma cells share certain characteristics with primary neurons, particularly in their highly elaborate axon structures, which make them suitable for *in vitro* studies simulating primary neuron attachment and proliferation[44]. These cells are also commonly employed in research on neurodegeneration and neuro-regeneration diseases[45–48]. The Neuroquantify models offer precise analysis of neurite structures, including measurements of neurite length and the distribution of orientations. The models allow users to analyze complex neuronal networks within seconds. Importantly, our large model achieves performance levels comparable to those of human experts.

Despite the promising segmentation results, certain aspects of our algorithms can be improved. Firstly, it is necessary to enhance the accuracy of neurite detection in the small model. This can be achieved by conducting additional training with a larger dataset, which would contribute to higher accuracy. Techniques like data augmentation can be employed to introduce more variations in the existing dataset during model training. Additionally, conducting further experimentation with hyperparameter tuning for different model parameters can help identify the most optimal configuration. Secondly, a limitation of our models is that the U-Net model has been trained specially using our dataset. As such, its performance for segmenting images captured with different types of microscopes or at varying magnification remains unverified. To make NeuroQuantify more universally applicable in diverse capture environments, it is imperative to gather additional datasets encompassing various imaging conditions.

While our primary objective has been to develop an algorithm for quick and efficient analysis of neuronal networks in neuroblastoma cells, NeuroQuantify also offers the possibility to analyze other cell types, such as PC12 cells (as depicted in **Figure S3**). Overall, NeuroQuantify demonstrates its potential for quantitatively assessing neuron cells and their associated networks. It exhibits a high level of effectiveness in segmenting cells and neurites within intricate structures, while also providing accurate quantitative measurements of their length and orientations.

## 5. Conclusion

In summary, we have introduced NeuroQuantify, a deep learning model for detecting and segmenting cells and neurites in phase contrast microscopy images, without the need for fluorescence labelling. NeuroQuantify can analyze the images within a few seconds to provide quantitative information about cell numbers, neurite numbers, neurite lengths, and neurite orientation distribution. These functionalities are useful for biological research requiring assessments of neuronal network development. In the future, we plan to implement and evaluate NeuroQuantify's performance in three-dimensional images and to improve the neurite detection accuracy to extend the applicability of this software.

**Author Contributions**

J.K.S.P. and K.M.D. conceived and, with P.C., supervised the project. Y.J.Z., T.Z., and C.W. developed deep learning models, algorithms, and user interface. A.S. labeled some of the images and conducted a proof of concept for U-Net. K.M.D., Y.J.Z., and A.S. analyzed data and validated the models. K.M.D., Y.J.Z., T.Z., C.W., P.C., and J.K.S.P. co-wrote the manuscript. All authors contributed to the article and approved the submitted version.

**Acknowledgments**

The authors would like to thank Prof. Dr. Tome Kosteski from University of Toronto, Dr. Andreas Marek from Max Planck Computing and Data Facility, and Dr. Michael G. K. Brunk from Max Planck of Microstructure Physics for helpful discussions. This work was supported by the Max Planck Society.

**Conflict of Interest**

The authors declare no conflict of interest